\documentclass[aps,prl,twocolumn,groupedaddress]{revtex4}
\begin{document}

\title{Scaling Limit of Vicious Walkers,
Gaussian Random Matrix Ensembles, \\
and Dyson Brownian Motions}

\author{Makoto Katori}
\email[]{katori@phys.chuo-u.ac.jp}
\altaffiliation{On leave from Department of Physics,
Faculty of Science and Engineering,
Chuo University, Kasuga, Bunkyo-ku, Tokyo 112-8551, Japan}
\affiliation{
University of Oxford, 
Department of Physics--Theoretical Physics, 
1 Keble Road, Oxford OX1 3NP, United Kingdom}
\author{Hideki Tanemura}
\email[]{tanemura@math.s.chiba-u.ac.jp }
\affiliation{
Department of Mathematics and Informatics,
Faculty of Science, Chiba University, 1-33 Yayoi-cho, Inage-ku,
Chiba 263-8522, Japan}

\date{\today}

\begin{abstract}
We study systems of interacting Brownian particles in
one dimension constructed as the diffusion scaling limits of
Fisher's vicious walk models. We define two types of 
nonintersecting Brownian motions, in which we impose
no condition (resp. nonintersecting condition forever)
in the future for the first-type (resp. second-type).
It is shown that, when all particles start from the origin, 
their positions at time 1 in the first-type (resp. 
at time 1/2 in the second-type) process are identically 
distributed with the eigenvalues of 
Gaussian orthogonal (resp.unitary) random matrices. 
The second-type process is described by the stochastic 
differential equations of the Dyson-type Brownian motions 
with repulsive two-body forces proportional 
to the inverse of distances. The present study demonstrates that
the spatio-temporal coarse graining of random walk models
with contact interactions can provide 
many-body systems with long-range interactions.\\
\nonumber PACS numbers: 05.40.-a, 02.50.Ey, 05.50.+q

\end{abstract}


\maketitle

The walks of independent random walkers, in which none of walkers
have met others in a given time period, are called
vicious walks. Since each random walk tends to a Brownian motion
in the diffusion scaling limit, an interacting system of
Brownian motions will be constructed as the scaling limit of
vicious walkers \cite{Fis84,HF84}.
In an earlier paper, we explicitly performed the scaling limit
of the vicious walkers in one dimension and derived the
spatial-distribution function for the nonintersecting
Brownian motions \cite{KT1}.

The purpose of this Letter is to study the scaling limit
of vicious walkers as a stochastic process, while the
previous paper reported its static 
properties at fixed times in order to clarify the relation
with the timeless theory of Gaussian random matrices 
\cite{Meh91}.
We first claim that the limit process is 
in general time-inhomogeneous, {\it i.e.} 
the transition probability depends on the time interval $T$,
in which the nonintersecting condition is imposed.
Then we study in detail two types of processes
in time $t$ defined by setting $T=t$ and 
$T \to \infty$, respectively. The former process is
related with the Gaussian orthogonal ensemble (GOE)
and the latter with the Gaussian unitary ensemble (GUE)
in random matrix theory.
This result demonstrates the fact that
in the nonintersecting processes distributions
at finite times depend on whether the nonintersecting
condition will be also imposed in the future or not.
We will derive the stochastic differential equations
for the latter type and show that the drift terms act
as the repulsive two-body forces proportional to the
inverse of distances between particles.
In other words, the scaling limit of vicious walkers can
realize the Dyson-type Brownian motion models (at
the inverse temperature $\beta=2$) \cite{Dys62}.
The Gaussian ensembles of random matrices can be regarded
as the thermodynamical equilibrium of Coulomb gas system
and that is the reason why Dyson introduced a one-dimensional
model of interacting Brownian particles with 
(two-dimensional) Coulomb repulsive potentials.
Here we should emphasize the fact that the vicious walkers,
however, have only contact repulsive interactions to
satisfy the nonintersecting condition. 
By taking the diffusion scaling
limit, we can extract global effective interactions
among walkers and they turn to be the long-ranged Coulomb-type
interactions \cite{Rem0}. Such emergence of long-range effects 
in macroscopic scales from systems having only short-ranged 
microscopic interactions is found only at critical points
in thermodynamical equilibrium systems, but it is a typical
phenomenon enjoyed by a variety of interacting particle systems
in far from equilibrium.

We also report the scaling limit
of vicious walkers with a wall restriction.
It will be shown that in this situation 
nonintersecting Brownian motions are discussed in
relation to the {\it chiral} (Laguerre-type) 
Gaussian ensembles of random matrices \cite{Meh91,Nag,Ver}. 

Let $\{R_{k}^{s_{j}}\}_{k \geq 0}, \
j \in \{1,2, \cdots, n\}$, be the $n$ independent symmetric
simple random walks on 
${\bf Z}=\{\cdots, -2, -1, 0, 1, 2, \cdots \}$ started
from $n$ positions, $2s_{1} < 2s_{2} < \cdots < 2s_{n},
s_{j} \in {\bf Z}$. 
We fix the time interval as a
positive even number $m$ and impose the nonintersecting 
condition
\begin{equation}
R_{k}^{s_{1}} < R_{k}^{s_{2}} < \cdots
< R_{k}^{s_{n}} \quad 1 \leq \forall k \leq m.
\label{eqn:nonint}
\end{equation}
The subset of all possible random walks, which satisfy
(\ref{eqn:nonint}), is the vicious walks up to time $m$.
For $2 e_{1} < 2 e_{2} < \cdots < 2e_{n},
e_{j} \in {\bf Z}$, let
$V_{n}(\{R_{k}^{s_{j}}\}_{k=0}^{m};
R_{m}^{s_{j}}=2e_{j})$ be the probability
that such vicious walks that the $n$ walkers arrive at
the positions $\{2 e_{j}\}_{j=1}^{n}$ at time $m$ are realized
in $2^{mn}$ random walks. 
We then consider, with a large value
$L >0$, the rescaled lattice ${\bf Z}/(\sqrt{L}/2)$,
where the unit length is $2/\sqrt{L}$, and let
$\tilde{R}_{k}^{x}$ be the symmetric simple random walk
started from $x$ on this rescaled lattice.
The following was proved
in the previous paper \cite{KT1}; 
for given $t > 0$ and
$0 \leq x_{1} < x_{2} < \cdots < x_{n},
y_{1} < y_{2} < \cdots < y_{n}$,
$
\lim_{L \to \infty}
V_{n}(\{ \tilde{R}_{k}^{x_{j}} \}_{k=0}^{Lt};
\tilde{R}_{Lt}^{x_{j}} \in [y_{j}, y_{j}+dy_{j}]) 
= f_{n}(t; \{y_{j}\}|\{x_{j}\}) d^{n} y
$
with
\begin{eqnarray}
&& f_{n}(t; \{y_{j}\}|\{x_{j}\}) 
= (2 \pi t)^{-n/2} 
s_{\xi(y)}\left(e^{x_{1}/t}, \cdots,
e^{x_{n}/t}\right) \nonumber\\
&& \qquad \times e^{- \sum
(x_{j}^2+y_{j}^2)/2t }
\prod_{1 \leq j < k \leq n}
(e^{x_{k}/t}-e^{x_{j}/t}), \quad
\label{eqn:fn}
\end{eqnarray}
where $s_{\lambda}(z_{1}, \cdots, z_{n})$ is the Schur
function \cite{FH91} and $\xi_{j}(y)=y_{n-j+1}-(n-j)$
\cite{Rem1}.
The function (\ref{eqn:fn}) shows the dependence on
the initial positions of particles $\{x_{j}\}$ and
the end-positions $\{y_{j}\}$ at time $t$
of the probability that $n$ independent one-dimensional
Brownian particles do not intersect up to time $t$.
Its summation over all end-positions is denoted
by 
$
{\cal N}_{n}(t; \{x_{j}\})
= \int_{y_{1}< \cdots < y_{n}} d^{n} y \
f_{n}(t; \{y_{j}\}| \{x_{j}\})
$
and its asymptote for $t \to \infty$ was calculated as
\begin{equation}
{\cal N}_{n}(t; \{x_{j}\})
\simeq (2 \pi)^{-n/2} t^{-\eta_{n}}
b_{n}(\{x_{j}\}) /c_{n}
\label{eqn:Nn}
\end{equation}
with $\eta_{n}=n(n-1)/4$,
$b_{n}(\{x_{j}\})=s_{\xi(x)}(1, \cdots, 1)
=\prod_{1 \leq j < k \leq n}(x_{k}-x_{j})/(k-j)$
and
$c_{n}=n ! (2^{3n/2} \prod_{j=1}^{n}\Gamma(j/2+1))^{-1}$,
where $\Gamma(z)$ is the gamma function \cite{KT1}.

Since the vicious walkers are defined by imposing
the nonintersecting condition (\ref{eqn:nonint})
up to a given time $m$, the process depends on the
choice of $m$. That is, the process is 
time-inhomogeneous.
This feature is inherited in the process obtained 
in the diffusion scaling limit as explained below.
Let $T > 0$ and consider the $n$ nonintersecting
Brownian motions in the time interval $[0,T]$.
For $0 \leq s < t \leq T, \
x_{1} < \cdots < x_{n}, y_{1} < \cdots < y_{n}$,
the transition probability densities from the
configuration $\{x_{j}\}$ at time $s$ to $\{y_{j}\}$
at $t$ is given by
\begin{equation}
g_{n}^{T}(s, \{x_{j}\}; t, \{y_{j}\}) =
\frac{f_{n}(t-s; \{y_{j}\}|\{x_{j}\}) 
{\cal N}_{n}(T-t; \{y_{j}\})}{{\cal N}_{n}(T-s; \{x_{j}\})},
\label{eqn:inhom}
\end{equation}
since the numerator in RHS gives the nonintersecting probability
for $[0,T]$ specified with the configurations $\{x_{j}\}$
and $\{y_{j}\}$ at times $s$ and $t$, respectively,
and the denominator gives the probability only specified
with $\{x_{j}\}$ at $s$,
where we have used the Markov property of the process.
The time-inhomogeneity is obvious, since RHS depends not only
$t-s$ but also $T-s$ and $T-t$. 

In this Letter we study two special cases of $T$ ;
case (A) $T=t$ and case (B) $T \to \infty$.
We will show that there is an interesting correspondence
between these choices of $T$ and the Gaussian ensembles
of random matrices \cite{Meh91}. 
In order to see it we consider
the limit $x_{j} \to 0 \ (1 \leq \forall j \leq n)$.

{\it Case} (A) $T=t$. 
Since $\lim_{t \to 0} 
f_{n}(t; \{y_{j}\}|\{x_{j}\})=\prod_{j=1}^{n} 
\delta(x_{j}-y_{j})$
with Dirac's delta functions, 
${\cal N}_{n}(0; \{x_{j}\})=1$ for any $\{x_{j}\}$,
and setting $T=t$ makes (\ref{eqn:inhom}) depend
only on $t-s$ \cite{Rem1-5}.
For $t > 0$ and $|x| \equiv \sum_{j=1}^{n} |x_{j}| \ll 1$
\begin{eqnarray}
&&{\cal N}_{n}(t; \{x_{j}\})
=(2 \pi t)^{-n/2} \prod_{1 \leq j < k \leq n}
(e^{x_{k}/t}-e^{x_{j}/t}) \nonumber\\
&& \quad \times \int_{y_{1} < \cdots < y_{n}} d^{n} y \
s_{\xi(y)}(1, \cdots, 1) 
e^{-\sum y_{j}^2/2t} \left( 1+ {\cal O}(|x|) \right)
\nonumber\\
&& = \frac{t^{n(n-1)/4}}{(2 \pi)^{n/2} c_{n}}
\prod_{1 \leq j < k \leq n} 
\frac{e^{x_{k}/t}-e^{x_{j}/t}}{k-j} \
\left( 1+ {\cal O}(|x|) \right).
\nonumber
\end{eqnarray}
Then we have
$$
g_{n}^{t}(0, \{0\}; t, \{y_{j}\})
= c_{n}t^{-\zeta_{n}} e^{-\sum y_{j}^2/2t}
\prod_{1 \leq j < k \leq n} (y_{k}-y_{j})
$$
with $\zeta_{n}=n(n+1)/4$ 
for $y_{1} < \cdots < y_{n}$.
If we set $t=1$ and assume $y_{1} < \cdots < y_{n}$,
then
$$
g_{n}^{1}(0, \{0\}; 1, \{y_{j}\})
= n ! \ g_{n}^{{\rm GOE}}(\{y_{i}\}),
$$
where $g_{n}^{{\rm GOE}}(\{y_{i}\})$ is the
probability density of eigenvalues
of GOE random matrices \cite{Rem2}.

{\it Case} (B) $T \to \infty$.
Let
$$
p_{n}(s, \{x_{j}\}; t, \{y_{j}\}) \equiv
\lim_{T \to \infty} 
g_{n}^{T}(s, \{x_{j}\}; t, \{y_{j}\}).
$$
By virtue of (\ref{eqn:Nn}) we can determine the
explicit form for any initial configuration $\{x_{j}\}$
in this case as
\begin{equation}
p_{n}(0, \{x_{j}\}; t, \{y_{j}\}) =
\frac{h_{n}(\{y_{j}\})}{h_{n}(\{x_{j}\})}
f_{n}(t; \{y_{j}\}|\{x_{j}\}),
\label{eqn:pn}
\end{equation}
with the Vandermonde determinant
$
h_{n}(\{x_{j}\})=\prod_{1 \leq j < k \leq n}
(x_{k}-x_{j}).
$
In particular, if we take the limit $x_{j} \to 0 
\ (1 \leq \forall j \leq n)$, we have
\begin{equation}
p_{n}(0, \{ 0 \}; t, \{y_{j}\})
= c^{\prime}_{n}
t^{-\zeta^{\prime}_{n}} e^{-\sum y_{j}^{2}/2t}
\prod_{1 \leq j < k \leq n}
(y_{k}-y_{j})^{2},
\label{eqn:pn0}
\end{equation}
with $\zeta^{\prime}_{n}=n^2/2$ and
$c^{\prime}_{n}=((2 \pi)^{n/2} \prod_{j=1}^{n-1} j!)^{-1}.$
In this case we set $t=1/2$ and 
assume $y_{1} < \cdots < y_{n}$. Then 
$$
p_{n}(0, \{0\}; 1/2, \{y_{j}\})
= n ! \ g_{n}^{{\rm GUE}}(\{y_{i}\}),
$$
where $g_{n}^{{\rm GUE}}(\{y_{i}\})$ is the
probability density of eigenvalues
of GUE random matrices.
In this case the nonintersecting condition will be imposed
forever ($ T \to \infty$), while in the case (A) there will 
be no condition in the future. The distributions
of particles at finite times depend on the condition
in the future.

By generalizing the calculation, which we did
in the case (A), for arbitrary $T$ and comparing the
result with (\ref{eqn:pn0}), we have
\begin{eqnarray}
&& 
\frac{g_{n}^{T}(0, \{0\}; t, \{y_{j}\})}
{p_{n}(0, \{0 \}; t, \{y_{j}\})} \nonumber\\
&=& \bar{c}_{n} T^{-n(n-1)/4} t^{n(n-1)/2}
\frac{{\cal N}_{n}(T-t; \{y_{j}\})}
{\prod_{1 \leq j < k \leq n}(y_{k}-y_{j})} \nonumber
\end{eqnarray}
with
$
\bar{c}_{n}=c_{n}/c^{\prime}_{n}=
\pi^{n/2} 2^{-n} 
\prod_{j=1}^{n} \Gamma(j+1)/\Gamma(j/2+1).
$
This equality can be regarded as the
multi-variable generalization of
Imhof's relation between the probability
distributions of Brownian meander
and the Bessel process \cite{Imh84}.

In the case (B) we have obtained the explicit expression
(\ref{eqn:pn}) for arbitrary $\{x_{j}\}$ and we can
derive a system of stochastic differential equations for 
the process. Using it we will explain why we have the GUE
distribution at time $t=1/2$. Let 
$$
a^{k}(\{x_{j}\})=\sum_{j=1; j \not= k}^{n}
\frac{1}{x_{k}-x_{j}} \quad
\mbox{for} \ k=1,2, \cdots n.
$$
It is easy to verify that
\begin{eqnarray}
\label{eqn:formula1}
&& a^{k}(\{x_{j}\})
=  \partial_{k} \log h_{n}(\{x_{j}\}), \\
\label{eqn:formula2}
&& \sum_{k=1}^{n}\left[ 
\partial_{k} a^{k}(\{x_{j}\})
+\left( a^{k}(\{x_{j}\}) \right)^{2} \right] =0,
\end{eqnarray}
where $\partial_{k}=\partial/\partial x_{k}$.
Using these equalities, we will prove shortly that 
$p_{n}(0, \{x_{j}\}; t, \{y_{j}\})$
solves the diffusion equation
\begin{equation}
\frac{\partial}{\partial t} u(t; \{x_{j}\}) =
\frac{1}{2} {\mit\Delta} u(t; \{x_{j}\})
+ \sum_{k=1}^{n} a^{k}(\{x_{j}\}) 
\partial_{k} u(t; \{x_{j}\}),
\label{eqn:diff}
\end{equation}
where ${\mit\Delta}=\sum_{k=1}^{n} \partial_{k}^{2}$.
This implies that the process defined 
in the case (B) is the system of $n$ 
particles with positions $x_{1}(t), x_{2}(t), \cdots, x_{n}(t)$
at time $t$ on the real axis, whose time evolution is governed
by the stochastic differential equations
\begin{equation}
dx_{k}(t)=a^{k}(\{x_{j}(t)\})dt + dB_{k}(t), \quad
k=1,2, \cdots, n,
\label{eqn:sde1}
\end{equation}
where $\{B_{k}(t)\}_{k=1}^{n}$ are the independent standard
Brownian motions;
$B_{j}(0)=0, \langle B_{j}(t) \rangle \equiv 0$ and
$\langle (B_{j}(t)-B_{j}(s))(B_{k}(t)-B_{k}(s)) \rangle
= |t-s| \delta_{jk}$.
Because of the scaling property of Brownian motion,
$\sqrt{\kappa} B_{j}(t)$ is equal to $B_{j}(\kappa t)$
in distribution. Then, if we set $t=2t^{\prime}$
and write $x_{k}(t)=\tilde{x}_{k}(t^{\prime})$,
(\ref{eqn:sde1}) is equivalent with the $\alpha=0,
\beta=2$ case of the equations
$
d\tilde{x}_{k}(t^{\prime})
=-\beta \partial_{k} W^{\alpha}(\{\tilde{x}_{j}(t^{\prime})\}) 
dt^{\prime} + \sqrt{2} dB_{k}(t^{\prime}), k=1,2, \cdots, n,
$
with
$
W^{\alpha}(\{x_{j}\})= \alpha \sum_{j=1}^{n}x_{j}^{2}/2
- \sum_{1 \leq j < k \leq n} \log(x_{k}-x_{j}).
$
When $\alpha=1$, they are known as the stochastic differential
equations for the Dyson Brownian motions at the
inverse temperature $\beta=1/k T$ and the stationary 
distribution $\propto \exp(-\beta W^{1}(\{x_{j}\})$
\cite{Dys62,Meh91}.
If $\alpha=0$, the factor $\exp(-\beta \sum x_{j}^2/2)$
is replaced by $\exp(-\sum \tilde{x}_{j}^2/4 t^{\prime})$
for finite $t^{\prime}$ and thus when $t^{\prime}=1/(2\beta)$
we may have the Gaussian distribution with $\beta$.
Setting $\beta=2$ gives $t=2t^{\prime}=1/2$.
It should be noted that the system of diffusion equations 
describing the Dyson Brownian motions with $\beta=2$ 
can be mapped to the free fermion model \cite{Rem0,BF97}.

Now we prove that (\ref{eqn:pn}) satisfies (\ref{eqn:diff}).
First we remark that \cite{KT1}
$$
f_{n}(t; \{y_{j}\}|\{x_{j}\})
= \det_{1 \leq j, k \leq n}
\left( (2 \pi t)^{-1/2} e^{-(x_{k}-y_{j})^2/2t} \right).
$$
That is, $f_{n}$ is a finite summation of the products of 
Gaussian kernels and thus it satisfies the diffusion equation
\cite{For89}.
Therefore
$$
\frac{\partial}{\partial t} p_{n}(t; \{y_{j}\}|\{x_{j}\})
= \frac{1}{2} \frac{h_{n}(\{y_{j}\})}{h_{n}(\{x_{j}\})}
 {\mit\Delta} f_{n}(t; \{y_{j}\}|\{x_{j}\}).
$$
Then we can find that, if $\{a^{k}(\{x_{j}\})\}$ satisfy
the equations
\begin{eqnarray}
&&\sum_{k=1}^{n} a^{k}(\{x_{j}\})
\frac{1}{h_{n}(\{x_{j}\})}
\partial_{k} f_{n}(t; \{y_{j}\}|\{x_{j}\}) \nonumber\\
&& \
= - \sum_{k=1}^{n} \partial_{k}
(1/h_{n}(\{x_{j}\}))
\partial_{k} f_{n}(t; \{y_{j}\}|\{x_{j}\}) \qquad
\label{eqn:eq1}
\end{eqnarray}
and
\begin{equation}
\sum_{k=1}^{n} a^{k}(\{x_{j}\}) 
\partial_{k}(1/h_{n}(\{x_{j}\})) 
= -\frac{1}{2} \sum_{k=1}^{n} 
\partial_{k}^{2} (1/h_{n}(\{x_{j}\})) ,
\label{eqn:eq2}
\end{equation}
the proof will be completed.
It is easy to see that (\ref{eqn:eq1}) is satisfied if
(\ref{eqn:formula1})
holds for all $k$.
Moreover, using (\ref{eqn:formula1}), we can reduce
(\ref{eqn:eq2}) to (\ref{eqn:formula2}).
Then the proof is completed.
This proof shows that the origin of the
long-ranged interactions (\ref{eqn:formula1}) is
the normalization factor $1/h_{n}(\{x_{j}\})$
in (\ref{eqn:pn}).

In order to derive the chiral versions of 
the Gaussian ensembles of random matrices 
\cite{Meh91,Nag,Ver} and of the Dyson Brownian motions, 
next we consider the vicious walker problem
with a wall restriction \cite{KGV00}.
We impose the condition
\begin{equation}
R_{k}^{s_{1}} \geq 0 \quad
1 \leq \forall k \leq m
\label{eqn:wall}
\end{equation}
in addition to (\ref{eqn:nonint}).
That is, there assumed to be a wall at the origin
and all walkers can walk only in the region 
$x \geq 0$. We write the
probability to realize the vicious walks which
satisfy these conditions (\ref{eqn:nonint}),
(\ref{eqn:wall}) and $\{R_{m}^{s_{j}}=2e_{j}\}_{j=1}^{n}$
as
$V_{n}^{+}( \{R_{k}^{s_{j}} \}_{k=0}^{m};
R_{m}^{s_{j}}=2e_{j})$.
By the Lindstr\"om-Gessel-Viennot theorem and
the reflection principle of random walks
\cite{KGV00}, we have
\begin{eqnarray}
&&V_{n}^{+}( \{R_{k}^{s_{j}} \}_{k=0}^{m};
R_{m}^{s_{j}}=2e_{j}) = 2^{-mn} \nonumber\\
&& \times \det_{1 \leq j, k \leq n} 
\left(
{ m \choose \frac{m}{2} +s_{k}-e_{j}} 
- { m \choose \frac{m}{2}+s_{k}+(e_{j}+1) }
\right).
\nonumber
\end{eqnarray}
Following the same calculation as was done in \cite{KT1},
we will obtain, for $t > 0,
0 \leq x_{1} < \cdots < x_{n}, \
0 \leq y_{1} < \cdots < y_{n}$, 
$ \lim_{L \to \infty}
V_{n}^{+}(\{ \tilde{R}_{k}^{x_{j}} 
\}_{k=0}^{Lt};
\tilde{R}_{Lt}^{x_{j}} \in [y_{j}, y_{j}+dy_{j}]) 
= f_{n}^{+}(t; \{y_{j}\}|\{x_{j}\}) d^{n} y,
$
with
\begin{eqnarray}
&& f_{n}^{+}(t; \{y_{j}\}|\{x_{j}\}) = (2 \pi t)^{-n/2} 
sp_{\xi^{+}(x)}\left(e^{y_{1}/t}, \cdots,
e^{y_{n}/t}\right)  \nonumber\\
&& \quad \times e^{- \sum
(x_{j}^2+y_{j}^2)/2t } \
\prod_{j=1}^{n}(e^{y_{j}/t}-e^{-y_{j}/t}) \nonumber\\
&& \quad \times \prod_{1 \leq j < k \leq n}
\{ (e^{y_{k}/t}-e^{y_{j}/t})(e^{(y_{j}+y_{k})/t}-1)\},
\nonumber
\end{eqnarray}
where $sp_{\lambda}(z_{1}, \cdots, z_{n})$ is the character
of the irreducible representation specified by
a partition $\lambda=(\lambda_{1}, \cdots, \lambda_{n})$
of the symplectic Lie algebra (see, for example,
Lectures 6 and 24 in \cite{FH91}) given by
$
\det(z_{j}^{\lambda_{k}+n-k+1}-z_{j}^{-(\lambda_{k}+n-k+1)})/
\det(z_{j}^{n-k+1}-z_{j}^{-(n-k+1)}),
$
and we set
$\xi^{+}(x)=(\xi^{+}_{1}(x), \cdots, \xi^{+}_{n}(x))$
with $\xi^{+}_{j}(x)=x_{n-j+1}-(n-j+1)$.
It is known that
$
sp_{\lambda}(1, \cdots, 1)=
\prod_{1 \leq j < k \leq n} 
(\ell_{k}^2-\ell_{j}^{2})/(m_{k}^{2}-m_{j}^{2})
\prod_{j=1}^{n} \ell_{j}/m_{j}
$
with
$
\ell_{j}=\lambda_{j}+n-j+1, m_{j}=n-j+1$ \cite{FH91}
and 
\begin{eqnarray}
&& \int d^{n} u \
e^{- \sum u_{j}^{2}/2}
\prod_{1 \leq j < k \leq n}
|u_{k}^{2}-u_{j}^{2}|^{2 \gamma}
\prod_{j=1}^{n} |u_{j}|^{2 \alpha-1} \nonumber\\
&& \ = 2^{\alpha n + \gamma n (n-1)}
\prod_{j=1}^{n} \frac{\Gamma(1+j \gamma) 
\Gamma(\alpha+\gamma(j-1))}{\Gamma(1+\gamma)}
\nonumber
\end{eqnarray}
for ${\rm Re} \, \alpha >0$ 
((17.6.6) on page 354 in \cite{Meh91}).
Then we can obtain the asymptote of
${\cal N}_{n}^{+}(t; \{x_{j}\})
= \int_{0 \leq y_{1} < \cdots < y_{n}} d^{n}y 
 f_{n}^{+}(t; \{y_{j}\}|\{x_{j}\}) $
for
$t \gg 1$ as
\begin{eqnarray}
&&{\cal N}_{n}^{+}(t; \{x_{j}\}) 
= t^{-n(2n+1)/2}(1+{\cal O}(1/t)) \
\frac{b_{n}^{+}(\{x_{j}\})}{(2 \pi)^{n/2} n!} \nonumber\\
&& \quad \times\int d^{n}y \
e^{-\sum y_{j}^{2}/2t}
\prod_{1 \leq j < k \leq n} |y_{k}^{2}-y_{j}^{2}|
\prod_{j=1}^{n} |y_{j}| \nonumber\\
&& \simeq t^{-\eta_{n}^{+}} b_{n}^{+}(\{x_{j}\}) 
\frac{2^{n(n+2)/2}}{\pi^{n} n!} \ \prod_{j=1}^{n} 
\Gamma\left(1+\frac{j}{2}\right) 
\Gamma\left(1+\frac{j-1}{2}\right)
\nonumber
\end{eqnarray}
with $\eta_{n}^{+}=n^2/2$, where 
$b_{n}^{+}(\{x_{j}\}) = \prod_{1 \leq j < k \leq n}
(x_{k}^{2}-x_{j}^{2})/(k^{2}-j^{2})
\prod_{j=1}^{n} x_{j}/j$.

Two types of diffusion processes can be 
defined as in the cases without a wall.
In the case (A) we will conclude that at time $t=1$ 
the positions of the nonintersecting Brownian particles
all started from the origin with a wall restriction 
is identically distributed with the positive eigenvalues of 
random matrices in the chiral GOE.
In the case (B) the transition probability
density is given as
$$
p_{n}^{+}(0, \{x_{j}\}; t, \{y_{j}\})
= \frac{h_{n}^{+}(\{y_{j}\})}{h_{n}^{+}(\{x_{j}\})}
f_{n}^{+}(t; \{y_{j}\}|\{x_{j}\})
$$
with
$h_{n}^{+}(\{x_{j}\})=\prod_{1 \leq j < k \leq n}
(x_{k}^{2}-x_{j}^{2}) \prod_{j=1}^{n} x_{j}$.
We can prove that $p_{n}^{+}(0, \{x_{j}\}; t, \{y_{j}\})$
satisfies the diffusion equation (\ref{eqn:diff}) if
$a^{k}(\{x_{j}\})$ is replaced by
$
a^{k+}(\{x_{j}\}) = \sum_{j=1; j \not= k}^{n}
2x_{k}/(x_{k}^{2}-x_{j}^{2})+ 1/x_{k}.
$
When $\{x_{j}\}=\{0\}$, the distribution at time $t=1/2$
may be equal to the distribution of the positive
eigenvalues of random matrices in the chiral GUE.

In summary, we have studied the diffusion processes derived as 
the scaling limits of vicious walkers in one dimension
with and without a wall.
Two types of interacting Brownian particles were
defined depending on the situations in the future and
interesting correspondence between these processes
and the Gaussian ensembles of random matrices
was discussed.
The systems of 
stochastic differential equations for the second-type
processes were determined, which can be 
identified with the Dyson Brownian motion models
at $\beta=2$ with appropriate change of time-scales. 
Further study of general time-inhomogeneous
processes determined by (\ref{eqn:inhom}) will be
an interesting future problem.

The authors acknowledge useful discussions with
T. Fukui, P. Forrester, T. Nagao and J. Cardy.


\begin{thebibliography}{99}
\bibitem{Rem1}
This is equal to the function defined by the
RHS of Equation (5) in \cite{KT1}, though
here we exchange the roles of $x_{j}$'s and
$y_{j}$'s.
\bibitem{Rem1-5}
It should be remarked that, however, 
$g_{n}^{t}(s, \{x_{j}\}; t, \{y_{j}\})$
does not satisfy the Chapman-Kolmogorov equation, while
$p_{n}(s, \{x_{j}\}; t, \{y_{j}\})$ given below
satisfies it.
\bibitem{Rem2}
This equivalence was proved in \cite{KT1}
by using the scaling property of Brownian motion
and (\ref{eqn:Nn}).
Here we give another proof using the expression 
(\ref{eqn:fn}).
\bibitem{Imh84}
J. -P. Imhof,
J. Appl. Prob. {\bf 21}, 500 (1984).
When $n=2$, we can consider the probability distributions
for one variable $y=y_{2}-y_{1}$.
In this case $g_{2}^{T}$ with $T=1$ and $p_{2}$ may be
related with the distributions of the Brownian meander 
and the Bessel process, respectively.
\end{thebibliography}

\end{document}